\documentclass[a4paper,12pt]{article}
\usepackage{verbatim} 
\usepackage{jheppub} 

\usepackage[T1]{fontenc} 

\usepackage[percent]{overpic}
\usepackage{wrapfig}
\usepackage{tabu}
\usepackage{diagbox}

\usepackage{graphicx}
\usepackage{tikz}
\usepackage{import}
\usepackage{accents}
\usepackage{mathrsfs,amsmath,amssymb,slashed}
\usepackage{multirow,multicol}
\usepackage[percent]{overpic}
\usepackage{slashed}

\usepackage{wrapfig}
\usepackage{tabu}
\usepackage{diagbox}
\usepackage{mathrsfs,amsmath,amssymb,amsthm,amsfonts,tikz,graphicx,accents,hyperref,color}
\usepackage{leftidx}
\usepackage{import}
\usetikzlibrary{decorations.pathmorphing}
\DeclareFontFamily{OT1}{rsfs}{}

\DeclareFontShape{OT1}{rsfs}{m}{n}{ <-7> rsfs5 <7-10> rsfs7 <10->rsfs10}{} 

\DeclareMathAlphabet{\mycal}{OT1}{rsfs}{m}{n}

\newcommand{\cD}{\mathcal{D}}

\newcommand{\nn}{\nonumber}

\newcommand{\be}{\begin{equation}}
\newcommand{\ee}{\end{equation}}

\DeclareMathOperator{\extdm}{d}
\newcommand{\extd}{\extdm \!}

\title{\bf {Exotic  massive 3D gravities from truncation}}
\author[]{Hamid Reza Afshar$^{\, a,b}$,}
\author[]{Nihat Sadik Deger$^{\, c,d}$}
\affiliation[a]{\it School of Physics, Institute for Research in Fundamental
Sciences (IPM),\\ P.O.Box 19395-5531, Tehran, Iran}
\affiliation[b]{\it Institute for Theoretical Physics, TU Wien, Wiedner Hauptstr.~8, A-1040 Vienna, Austria
 }
 \affiliation[c]{Department of Mathematics, Bogazici University,
Bebek, 34342, Istanbul, Turkey}
\affiliation[d]{Erwin Schr\"odinger International Institute for Mathematics and Physics,\\ Boltzmanngasse 9, 1090 Vienna, Austria}
\emailAdd{afshar@ipm.ir}
\emailAdd{sadik.deger@boun.edu.tr}

\abstract{
We introduce a systematic way of constructing 3D exotic massive gravity theories in the first order formulation.
Our method is based on truncating a single degree of freedom in the parity odd gravity models found earlier \cite{Afshar:2014ffa} and supplementing it with appropriate potential terms such that the resulting models have well-defined metric equations but their Bianchi identities are satisfied only on-shell. Hence, they are `third way' consistent. We first re-derive two already
known exotic theories using our approach and then construct an extended exotic massive gravity model whose metric field equation is sixth order in derivatives.
We also explain how to check Bianchi identities using the first order formulation.}

\begin{document}
\maketitle

\section{Introduction}

Higher derivative interactions for gravity theories have been an interesting playground in various studies in cosmology, black hole physics and AdS/CFT duality. 
However, models whose equations of motion contain more than two derivatives are usually plagued with ghosts since a new massive mode arises which is tachyonic. 
In three dimensions (3D) one may play with the sign in front of the action by arguing that the massless mode does not propagate but this would spoil unitarity in the dual CFT picture as charges of the theory are calculated at infinity and contributed by the massless sector. This is usually referred as the bulk-boundary unitarity clash in 3D gravities \cite{Li:2008dq}.

Constructing gravity models with any number of derivatives in 3D is very systematic in the first order formalism where the Lorentz and the diffeomorphism invariances are manifest. The Lagrangian is a 3-form and should be written in a gauge invariant manner based on the Lorentz gauge group SO(1,2) with respect to the spin connection $\omega$ \cite{Afshar:2014ffa}. These models fit into the family of the `Chern-Simons-like' theories \cite{Hohm:2012vh,Bergshoeff:2014bia}. Here, in addition to the dreibein and the spin connection, we may also allow a number of auxiliary form fields if we want a gravity model with higher number of derivatives in the metric formulation. Provided that one can solve for these auxiliary fields algebraically in terms of the dreibein and its derivatives ($e_\mu{}^a,\partial e_\mu{}^a,\cdots,\partial^ne_\mu{}^a$),
 one can insert them back into the action or field equations and obtain their metric form. So, we can associate a weight $n$ to each of these fields. We denote those with even weights ($n=2I$) 
 as $f_I$ and those with odd weights ($n=2I+1$) as $h_I$. The dreibein $e$ and the spin connection $\omega$ correspond to $I=0$, see table \ref{t1}. All fields are Lorentz vector 1-forms
\begin{align}
    e=e^aJ_a\,,\qquad\quad\omega=\omega^aJ_a\,,\quad\qquad f_I=f_I^aJ_a\,,\quad\qquad h_I=h_I^aJ_a\,,
\end{align}
with $X^a=X_\mu^a \extd x^\mu$ being a generic one-form and $J^a$ being the generator of the 3D gauge Lorentz algebra, $[J_a,J_b]=\epsilon_{abc}J^c$. 
 This formulation by construction avoids scalar ghosts and leads to the right number of degrees of freedom for the (massive) graviton. 
\begin{table}[ht]
 \label{t1}
\begin{center}
  \begin{tabular}{l l l l  l l  l l  }
  Fields &  $e$ & $\omega$ & $f_1$ & $h_1$ & $f_2$ & $h_2$ & $\cdots$\\     \hline\hline
   Weight & 0 & 1 & 2 & 3 & 4 & 5 & $\cdots$\\
  \end{tabular}
 \caption{Form fields and their weights}
   \end{center}
\end{table}

There are two gauge invariant sectors namely, parity even theories with $2N+2$ derivatives and parity odd theories with $2N+3$ derivatives (where $N\geq0$) whose actions are
denoted as $S_{2N}$ and $S_{2N+1}$ respectively, see table \ref{t2}.
The systematic construction of these models was considered in \cite{Afshar:2014ffa}. Kinetic and potential terms in these actions with a definite parity are given as follows:
\begin{itemize}
    \item Parity even terms:
    \begin{align}\label{parityeven}
         \left\langle f_I\wedge\mathcal Dh_{J}\right\rangle\,,
         \qquad\left\langle f_I\wedge f_J\wedge f_K\right\rangle\,,
         \qquad \left\langle f_I\wedge h_J\wedge h_{K}\right\rangle\,.
    \end{align}
    The total weight in each of these terms should not exceed the number of derivatives which is $2N+2$ here. The cases of $N=0$ and $N=1$ correspond to Einstein-Hilbert gravity $S_0$ and new massive gravity (NMG) $S_2$ \cite{Bergshoeff:2009hq, Bergshoeff:2009aq} respectively.
    \item Parity odd terms:
    \begin{align}\label{parityodd}
        \left\langle h_I\wedge\mathcal Dh_{J}\right\rangle\,,
        \qquad\left\langle f_I\wedge\mathcal Df_{J}\right\rangle\,,
        \qquad\left\langle h_I\wedge h_J\wedge h_K\right\rangle\,,
        \qquad\left\langle f_I\wedge f_J\wedge h_{K}\right\rangle\,.
    \end{align}
    The total weight in each of the above terms should not exceed the number of derivatives which is $2N+3$. The $N=0$ case $S_1$, corresponds to 3D conformal  gravity \cite{Deser:1981wh,Horne:1988jf,Afshar:2011qw,Afshar:2013bla}.
\end{itemize}
Here, $\cD$ is the exterior covariant derivative with respect to the 3-dimensional Lorentz group {which has weight one} and $\langle\,\rangle$ indicates the appropriate contraction of Lorentz indices  {such that by dropping the wedge symbol we have; $\left\langle A\wedge B \right\rangle=A\cdot B$ and $\left\langle A\wedge B\wedge C \right\rangle=\tfrac12A\cdot[B,C]$}.

  As stated above, the guiding principle in this construction is simply to respect the weight in each term according to the number of derivatives in the model.\footnote{The spin connection $\omega\equiv h_0$ obviously can only appear {in Kinetic terms via the covariant derivative $\mathcal D$ and the Ricci 2-form $R$} and in potential terms as the Chern-Simons combination.} As was shown in \cite{Afshar:2014ffa}, both sectors of these models have a metric formulation at the level of their action.
Once we exhaust all possible gauge invariant 3-form terms (including the kinetic and the potential ones) at a given weight, this has the following consequences:

\begin{enumerate}
\item The torsion remains zero by which one can solve the spin connection as usual.
    \item The Bianchi identity is guaranteed to hold off-shell. 
    \item All auxiliary fields are solved algebraically since they appear linearly.
    \item The field equation and the action has the same parity.
\end{enumerate}
The second item above essentially implies that the theory will have an action in the metric formulation. Parity violating models can be constructed by combining these parity even and odd models, e.g. topologically massive gravity (TMG) \cite{Deser:1981wh} as $S_0+S_1$, general massive gravity (GMG) \cite{Bergshoeff:2009hq, Bergshoeff:2009aq} as $S_1+S_2$  and so on. 


There exists a modification of this first order construction of higher derivative gravity models such that among the four properties listed above, the first and the third property are retained while the second property holds only on-shell and the fourth one is relaxed. 
In this modification the assumption of preserving the weight in each potential/kinetic term of \eqref{parityeven} and \eqref{parityodd} is relaxed in such a way that the third property above still holds, the first property can be restored by a linear shift in the spin connection while the second property only holds on-shell. As a consequence there is no action in the metric formulation and the 4th property does not necessarily hold any more. This is referred to as the {\it third way consistency}. The first example of such a model was found in \cite{Bergshoeff:2014pca} and was called minimal massive gravity (MMG). It attracted a lot of attention since it offered a possible resolution to the bulk-boundary clash 
\cite{Arvanitakis:2014xna,Arvanitakis:2015oga,Bergshoeff:2015zga}. Another example was found in \cite{Ozkan:2018cxj} and was called exotic massive gravity (EMG) which has a parity odd first order action but parity even field equations. They were obtained by deforming the $S_1$ and $S_2$ actions respectively. Our goal in this paper is to show that exotic models, which we denote as ${\tilde S}_{2N}$, can be constructed via truncation of parity odd actions 
\begin{align}\label{exotictrun}
    S_{2N+1}\to \tilde{S}_{2N}\,,
\end{align}
in which the highest weight field $f_{N+1}$ in $ S_{2N+1}$ is identified with {a linear combination of} lower weight ones $f_N,\, f_{N-1},\cdots$ in a parity preserving manner. {One also needs to add irrelevant, i.e. weight violating, potential terms including $h_N^3$ in the action such that after a shift in the spin-connection  by $h_N$ (and possibly also with lower weight $h_i$'s), the theory is third way consistent.}\footnote{ See \cite{Alkac:2018eck} for construction of such models directly in the metric formulation.}
 In this top-down approach no guess work is involved {for the dynamical terms in ${\tilde S}_{2N}$ and the necessary potential terms are not difficult to figure out by requiring  the system to be third-way solvable}. 
 Hence, it 
is more systematic than the bottom-up approach employed earlier \cite{Bergshoeff:2014pca, Ozkan:2018cxj}, which is hard to generalize if one wants models with even higher order derivatives. These exotic models have the same number of derivatives and degrees of freedom as in $S_{2N}$, see table 2. 


\begin{table}[ht]
\begin{center}
  \begin{tabular}{ l  l l  l  l l}
       & & $S_{2N}$ & ${\tilde S}_{2N}$ & $S_{2N+1}$ \\
    \hline\hline
       parity &    & even & odd & odd \\
    \hline
       $\#$ deriv. &  &  $2N+2$ & $2N+2$ & $2N+3$ \\
    \hline
      $\#$ d.o.f. &  & $2N$ & $2N$ & $2N+1$\\
  \end{tabular}
         \caption{\label{t2}
Parity preserving models, their number of derivatives and local degrees of freedom, $N\geq0$. $S_1$ is special and does not obey the rule in number of degrees of freedom as it enjoys one extra conformal gauge symmetry and consequently has zero d.o.f.}
   \end{center}
\end{table}
The structure of this paper is as follows. In  section \ref{3rdway} we explain how to check Bianchi identities using the first order formulation which is easier than doing this at the metric 
level. In section \ref{EMG1} we re-derive known examples of exotic massive gravity models, that is ${\tilde S}_0$ \cite{Witten:1988hc} and ${\tilde S}_2$
\cite{Ozkan:2018cxj},
using our truncation idea. In section \ref{EMG2} we apply our approach to the $S_5$ action to construct the next order example of exotic massive gravity namely ${\tilde S}_4$. We show that it is third way consistent and give its metric field equation (\ref{metriceom}) which is of order six.
We conclude in section  \ref{conc} by indicating some future directions. In appendix \ref{mmgemg} we show that MMG \cite{Bergshoeff:2014pca} can be obtained from EMG \cite{Ozkan:2018cxj} using a parity violating truncation.




\section{Bianchi identities in the first order formulation}\label{3rdway}
 
For higher derivative gravity theories checking the Bianchi identity becomes quite complicated as number of derivatives increase. However, when a first order formulation is available, this computation 
is rather straightforward which  we would like to illustrate in this section.
 
 In the models constructed in \cite{Afshar:2014ffa}, field
 equations in the first order formulation can be 
ordered from that of the lowest degree  field, i.e. dreibein, to 
the highest degree (auxiliary) field such that at level $n$, the
degree $n$ field appears linearly as the unknown. Assuming invertibility of the dreibein, these equations can be solved one-by-one algebraically (with finite number of terms) until the last which becomes the metric field equation of the model.\footnote{Here, we only consider a frame formalism which leads to finite number of terms in the metric formulation. There are examples such as the multiple interacting frame fields \cite{Bergshoeff:2013xma,Afshar:2014dta} and the Born-infeld gravity \cite{Gullu:2010pc} in its frame form --- see the discussion session of \cite{Afshar:2014ffa} --- whose metric formulation leads to infinite number of terms.}   Checking 
the Bianchi identity in this formalism amounts to checking whether the covariant derivative of the 2-form appearing on the left hand side of the last equation is zero. In doing that, one is allowed to use all previous equations and their covariant derivatives which give several constraints. These intermediate constraints are however  trivially satisfied  in the 2nd order formulation of the theory. If the Bianchi identity is satisfied only after using the last equation 
itself, then we have a third way consistent model. Note that this means that the model does not have a covariant metric formulation as the Bianchi identity is just a consequence of the diffeomorphism invariance. A clear sign of such a model is to have a square of the highest degree form field appearing in the last first order equation.



\paragraph{(Exotic) Einstein gravity.}
To construct a gravity model in three dimensions in the first order formulation obviously we at least need the dreibein and the spin-connection. 
The number of degrees of freedom for this minimal set of fields is zero which is due to the fact that
the number of dynamical spatial components $e_i^a$ and $\omega_i^a$ is 12 and there are six diagonal gauge symmetries and six temporal components $e_0^a$ and $\omega_0^a$ as Lagrange multipliers. 

Einstein field equations in the presence of a cosmological constant is:
\begin{equation}\label{Einsteineq}
    G_{\mu\nu}+\Lambda_0g_{\mu\nu}=0\,,
\end{equation} 
which in 3-dimensions can be derived from the following first order field equations on the dreibein $e=(e_\mu{}^a \extd x^\mu) J_a$ and the spin-connection $\omega=(\omega_\mu{}^a \extd x^\mu) J_a$:
\begin{subequations}\label{EGeom}
\begin{align}
    &\cD e=0\,,\label{torsion1}\\
   & R-\tfrac{1}{2}\Lambda_0 [e,e]=0\label{Einsteom}\,.
\end{align}
\end{subequations}
Here $\cD e=\extd e+[\omega,e]\equiv T$ and $R\equiv\extd\omega+\tfrac12[\omega,\omega]$ with $\cD\equiv\extd+[\omega,\;]$ being  the exterior covariant derivative with respect to the SO(1,2) gauge field $\omega$. 
Interestingly, field equations \eqref{EGeom} can be integrated to the level of first order actions in two different ways;
\begin{align}
     S_0[e,\omega]&=-\frac{1}{\kappa^2}\int \langle e\wedge R -\frac{\Lambda_0}{3}e\wedge e\wedge e\rangle\,,\label{einstein}\\
    \tilde{S}_0[e,\omega]&=\frac{1}{2\kappa^2\mu}\left(S_{\text{\tiny LCS}}-\Lambda_0\int \langle e\wedge \cD e\rangle\right)\,,\label{exoticEH}
\end{align}
where $\mu$ is an arbitrary mass scale and
the Lorentz Chern-Simons action is given as;
\begin{align} \label{csaction}
    S_{\text{\tiny LCS}}[\omega]=\int\langle\omega\wedge\extd\omega+\tfrac23\omega\wedge\omega\wedge\omega\rangle\,.
\end{align} 
The action \eqref{einstein} describes the ordinary Einstein gravity whereas the gravity theory obtained from \eqref{exoticEH} is called as exotic 3D gravity. One should note that in this model the cosmological constant $\Lambda_0$ can not be set to zero as it guarantees the torsion constraint \eqref{torsion1}. The $S_0$ and $\tilde{S}_0$ both have Chern-Simons formulations based on the $so(2,2)$ algebra which can be written as the difference and the sum of two $sl(2,R)$ Chern-Simons theories, respectively \cite{Achucarro:1987vz,Witten:1988hc}. This is the consequence of the fact that the $so(2,2)$ algebra admits two non-degenerate bilinear forms \cite{Witten:1988hc,Townsend:2013ela}.

To check the Bianchi identity, we apply the covariant exterior derivative on the 
first equation \eqref{torsion1} and get the constraint
\begin{equation}
    0= \cD\cD e =  [R, e] \,,\label{BianchiEG1}
\end{equation}
which is satisfied in the second order formulation of the theory where the spin-connection is solved in terms of the dreibein from the torsion-zero constraint \eqref{torsion1}. Now applying the covariant exterior derivative on the left hand side of the field equation 
(\ref{Einsteom}) and again using the equation \eqref{torsion1},  we get
\begin{align}
    \cD (R-\tfrac{1}{2}\Lambda_0 [e,e]) = \cD R \label{BianchiEG2} \, ,
\end{align}
 which is identically zero. Note that in getting to this result we have not used the field equation (\ref{Einsteom}) itself.
This is of course not surprising as we know that the Einstein equation  \eqref{Einsteineq} can be obtained from the Einstein-Hilbert action.

\paragraph{Conformal Chern-Simons gravity (CSG).}
In the next level we can add a new Lie algebra valued 1-form field $f_1=\left(f_1{}^a_\mu\extd x^\mu\right)J_a$ whose weight is 2 and
write the following parity odd gauge invariant action,
\begin{align}\label{CSGaction}
   S_{\text{\tiny CSG}}= S_1[e,\omega,f_1]&=\frac{1}{2\kappa^2\mu}\left(S_{\text{\tiny LCS}}+2\int \langle f_1\wedge \cD e\rangle\right)\,, 
\end{align}
which is the first order formulation of conformal gravity in three dimensions which leads to a model that is third order in derivative in the metric formulation \cite{Horne:1988jf,Afshar:2011qw,Afshar:2013bla}.  
The field equations of \eqref{CSGaction} are
\begin{subequations}\label{CSGeom}
\begin{align}\label{CSGeom1}
\delta f_1\qquad& \cD e = 0\,,  \\
\delta \omega\qquad& R + [e,f_1]=0\,,\label{Schouten}\\
\delta e\qquad& \cD f_1=0  \,.\label{cotton}
\end{align}
\end{subequations}
Equation \eqref{CSGeom1} implies that there is no torsion and the field $f_1$ appears linearly in (\ref{Schouten})
which can be solved easily as 
\begin{equation}
    f_1^a = -S^{ab}e_b \label{seqn}\, , 
\end{equation}
where $S_{\mu \nu}=  R_{\mu\nu}-\tfrac{1}{4}Rg_{\mu \nu}$ is the Schouten tensor. Now using this in (\ref{cotton}) we get the metric field equation of the model as
\begin{equation}
    C_{\mu\nu}= 0 \, , \label{ceqn}
\end{equation}
where $C_{\mu \nu}= e^{-1} \epsilon_{(\mu|}{}^{\alpha\beta} \nabla_{\alpha} S_{\beta |\nu)}$ is the Cotton tensor which is  symmetric, traceless and divergence free. The last property
simply means that Bianchi identity is satisfied off-shell which is
a consequence of the fact that the field equation (\ref{ceqn})
can be derived from the gravitational  Chern-Simons action:
\begin{align}
    S=\frac{k}{4\pi}\int d^3x \epsilon^{\lambda\mu\nu}\Gamma^\sigma_{\lambda\rho}\left(\partial_\mu\Gamma^\rho_{\nu\sigma}+\tfrac23\Gamma^\rho_{\mu\tau}\Gamma^\tau_{\nu\sigma}\right)\,.
\end{align}
Now, we would like to show that the Bianchi identity of this model is satisfied off-shell using its first order formulation (\ref{CSGaction}). Applying the covariant derivative on the equation (\ref{CSGeom1}) and using \eqref{Schouten}
we get the constraint
\begin{eqnarray}
  0=\cD\cD e=[R,e]=-[[e,f_1],e]=e\cdot f_1 e \label{const1} \, ,
\end{eqnarray}
where the dot $\cdot$ in the last item indicates contraction of the Lorentz indices. Note that this constraint is satisfied in the metric formulation where $f_1$ is given by \eqref{seqn}. Now applying the covariant derivative on the field equation (\ref{cotton}) we get
\begin{equation}
    \cD\cD f_1=[R, f_1]=-[[e,f_1], f_1]= e\cdot f_1 f_1 \, ,
\end{equation}
which vanishes due to (\ref{const1}). 

\paragraph{Topologically massive gravity (TMG).}A natural diffeomorphism invariant theory in this series is the topologically massive gravity (TMG) \cite{Deser:1981wh} which is the sum of $S_0$ \eqref{einstein}
and $S_1$ \eqref{CSGaction}
and as a consequence, parity violating:
\begin{align} \label{tmg}
    S_{\text{\tiny TMG}}=S_0+S_1 \,.
\end{align}
Field equations are obtained as
\begin{subequations}\label{tmgeqn}
\begin{align}
\delta f_1 \qquad& \cD e = 0\,,  \label{TMGeom1}\\
\delta \omega\qquad& R + [e,f_1] =0\,,\label{TMGeom2}\\
\delta e\qquad& \tfrac{1}{\mu}\cD f_1-R+\tfrac{1}{2}\Lambda_0 [e,e]=0  \,.\label{TMGeom3}
\end{align}
\end{subequations}
Since equations \eqref{TMGeom1}-\eqref{TMGeom2} are identical with (\ref{CSGeom1})-(\ref{Schouten}) the constraint \eqref{const1} is also valid in this model which immediately implies that the Bianchi identity is satisfied off-shell as expected.

\paragraph{Minimal massive gravity (MMG).}
Until now we have considered terms  which have dimensions up to 3 in the action and observed that in all these models Bianchi identity is satisfied off-shell. Since we have a new field $f_1$ with mass dimension 2 in the game, we may try to  deform the TMG action \eqref{tmg} by an irrelevant dimension-4 term as follows:
\begin{align} \label{mmgaction}
    S_{\text{\tiny MMG}}=S_{ \text{\tiny TMG}} +\frac{\alpha_1}{\kappa^2\mu^2}\int \langle e\wedge f_1\wedge f_1\rangle \, , 
\end{align}
where $\alpha_1$ is a dimensionless parameter. This is a minimal extension of our third order-in-derivative TMG model and is not going to affect the number of local degrees of freedom. This model is called as minimal massive gravity (MMG) \cite{Bergshoeff:2014pca} 
with the following field equations;
\begin{subequations}
\begin{align}\label{TMGeom}
\delta f\qquad& \cD e + \tfrac{\alpha_1}{\mu}\, [e,f_1]= 0\,,  \\
\delta \omega\qquad& \tfrac{1}{\mu}\left(R +  [e, f_1]  \right)-\cD e=0\,,\\
\delta e\qquad& \tfrac{1}{\mu}\cD f_1+\tfrac{\alpha_1}{2\mu^2} [f_1,f_1]-R +\tfrac{1}{2}\Lambda_0 [e,e]=0  \label{TMG3} \,.
\end{align}
\end{subequations}

The theory is not torsion free but we can make it so by making a parity violating shift\footnote{This is legitimate as TMG itself is parity violating.}:
\begin{align}
    \omega\to\omega -\frac{\alpha_1}{\mu} f_1 \, , 
\end{align}
which leads to;
\begin{align}
& \cD e  \rightarrow  \cD e - \frac{\alpha_1}{\mu} [e,f_1]\,,\\
&\cD f_1  \rightarrow \cD f_1 - \frac{\alpha_1}{\mu} [f_1,f_1] \,,\\
&R  \rightarrow  R - \frac{\alpha_1}{\mu} \cD f_1 + \frac{\alpha^2_1}{2\mu^2}[f_1,f_1] \,.
\end{align}
Consequently, field equations (\ref{TMGeom})-(\ref{TMG3}) transform as;
\begin{subequations}
\begin{align}
& \cD e = 0\,,  \label{m1}\\
& R +(1+\alpha_1)^2 [e,f_1 ] +\tfrac{\alpha_1}{2}\Lambda_0 [e,e]=0\,,\label{m2} \\
    &\tfrac{1+\alpha_1}{\mu}\cD f_1-\tfrac{\alpha_1(1+\alpha_1)}{2\mu^2}[f_1,f_1]-R+\tfrac{1}{2}\Lambda_0 [e,e]=0\,. \label{m3}
\end{align}
\end{subequations}
Note that, in comparison to TMG equations \eqref{tmgeqn} here we have the $f_1^2$ term in \eqref{m3} as a new ingredient which will spoil the Bianchi identity as we will show now.
The constraint \eqref{const1} is still valid for this model.
Now, taking the covariant derivative on the left hand side of the field equation (\ref{m3}) we find that it is proportional to 
\begin{equation}
    f_1\cD f_1 
\end{equation}
which is not identically zero but it vanishes if we 
replace $\cD f_1$ from the field equation \eqref{m3}:
\begin{align}
f_1\cD f_1 =[f_1,\cD f_1]\approx \frac{\alpha_1}{2\mu}\,[ f_1,[f_1,f_1]]=0\,.
    \end{align}
where we used the constraint \eqref{const1} again and the symbol $\approx$ means on-shell. Therefore, MMG is a third way consistent theory.

\section{Exotic 3D massive gravities from truncation}\label{EMG1}
In this section we explain how exotic gravity models 
${\tilde S}_0$ \eqref{einstein}
and ${\tilde S}_2$ \cite{Ozkan:2018cxj}
can be obtained starting 
from parity 
odd actions $S_1$ \eqref{CSGaction}
and $S_3$ found in \cite{Afshar:2014ffa}
respectively, by truncating the highest degree auxiliary field
in a parity preserving manner.

\paragraph{Conformal to Exotic gravity.}Here we make a simple but important observation which is going to be the basis of our construction of exotic models. The exotic action  $\tilde{S}_0$ in \eqref{exoticEH} can be obtained from conformal gravity action $S_1$ in \eqref{CSGaction} through a truncation of the extra field $f_1$ as follows;
\begin{align}\label{S0tilde}
	{\tilde S}_0 [e,\,\omega]= S_1[e,\,\omega,\,f_1\to -\tfrac{\Lambda_0}{2}e]\,.
\end{align}
This truncation obviously preserves the parity as it identifies an even form field $f_1$ (with weight 2) with the Dreibein $e$ (with weight 0). 
\paragraph{Exotic massive gravity (EMG).}
In order to construct ${\tilde S}_2$, i.e. 
EMG \cite{Ozkan:2018cxj}, we need the parity odd action $S_3$ that was first introduced in \cite{Afshar:2014ffa}:
\begin{align}\label{CSMG}
    S_3[e,\omega,f_1,h_1,f_2]=\frac{1}{\kappa^2\mu^3}\int\left\langle e\cD f_2+h_1\left(R+ef_1\right)+\tfrac{\alpha}{2} f_1\cD f_1\right\rangle + S_1\,.
\end{align}
Here $\alpha$ is a free parameter and from now on we will not put $\wedge$ between forms for simplicity.
It is also possible to extend 
this 5th order action by adding $S_0$ \eqref{einstein}
and $S_2$ in a parity violating manner. This model, which has three degrees of freedom, has a metric formulation and the Bianchi identity is satisfied off-shell as a consequence --- see appendix B in \cite{Afshar:2014ffa}.

Having introduced $S_3$, which is the next to leading parity odd action with metric formulation, we may now ask if there exists a 4th order-in-derivative parity odd model $\tilde{S}_2$
which has parity even field equations as in NMG \cite{Bergshoeff:2009hq}. To construct this model we start 
with $S_3$ given in \eqref{CSMG} and truncate a single degree of freedom by identifying the highest weight 1-form $f_2$ with lower even weight ones as $f_2=\mu^2\left(f_1-\tfrac{\Lambda_0}{2}  e\right)$
where $\mu$ and $\Lambda_0$ are some mass parameter constants. It turns out that in order to be able to solve for the spin connection $\omega$ and auxiliary fields $f_1$ and $h_1$ one after the other, we should further set $\alpha=0$ and also minimally deform the theory by the irrelevant $h_1^3$ term {so that at the end the theory is third way consistent}. 
So, we get
\begin{align}\label{S2tilde}
	{\tilde S}_2 [e,\,\omega,f_1,h_1]&= S_3[f_2\to \mu^2\left(f_1-\tfrac{\Lambda_0}{2}  e\right),\alpha=0]+\frac{2\zeta}{\kappa^2\mu^7}\int\left\langle h_1^3\right\rangle \\
	&=\frac{1}{\kappa^2\mu^3}\int  \left \langle \mu^2(f_1-\tfrac{\Lambda_0}{2}\,  e)\cD e+  h_1(R+ef_1+\zeta\,\mu^{-4} h_1^2)\right\rangle \, , \nn
\end{align}
where we omitted $S_\text{\tiny{LCS}}$ given in \eqref{csaction}
since it will not affect the following discussion.
The field 
equations are given as;
\begin{subequations}
\begin{align}
    \delta f_1\qquad&\cD e+\tfrac{1}{\mu^2}\,[e,h_1]=0\,,\\
    \delta h_1\qquad&R+[e,f_1]+\tfrac{3\zeta}{\mu^4}\,[h_1,h_1]=0\,,\\
    \delta e\qquad&\cD f_1+\tfrac{1}{\mu^2}\,[h_1,f_1]+\tfrac{\Lambda_0}{\mu^2}\, [e,h_1]=0\,,\\
    \delta \omega\qquad&\tfrac{1}{\mu^2}\cD h_1
    -\tfrac{\Lambda_0}{2}[e,e]=0\,.
\end{align}
\end{subequations}
After doing the shift $\omega\to\omega-\tfrac{1}{\mu^2}h_1$
so that the theory is torsion free, we see that choosing 
 $\zeta=\frac{1}{6}$ we get
\begin{subequations}
\begin{align}
    \delta f_1\qquad&\cD e=0\,,\\
    \delta h_1\qquad&R+[e,f_1]-\tfrac{\Lambda_0}{2}[e,e]=0\,,
    \label{f2}\\
    \delta e\qquad&\cD f_1+\tfrac{\Lambda_0}{\mu^2}\, [e,h_1]=0\,, \label{f23}\\
    \delta \omega\qquad&\tfrac{1}{\mu^2}\left(\cD h_1-\tfrac{1}{\mu^2}[h_1,h_1]\right)-\tfrac{\Lambda_0}{2}[e,e]=0\,.
\label{f4}
\end{align}
\end{subequations}
In the last equation the presence of the $h_1^2$ term signals 
that Bianchi identity is not satisfied off-shell and 
indeed the system is third way consistent. To see this, note that in addition to the constraint \eqref{const1}, here from \eqref{f2} and \eqref{f23} we also have
\begin{equation}
     e\cdot h_1 \,e= 0\,,
\end{equation}
which is trivially satisfied in the 2nd order formulation where $h_1$ is solved from \eqref{f23} as
\begin{align}
      h_1^a=C^{ab}e_b\,.
\end{align}
Now applying the  covariant derivative to the left hand side of  \eqref{f4} we see that it is zero only if we use the equation \eqref{f4}:
\begin{equation}
    h_1\cD h_1= [h_1,\cD h_1]\approx \tfrac{1}{\mu^2} [h_1,[h_1,h_1]]=0\,.
\end{equation}
The full exotic massive gravity theory is obtained after
adding $S_\text{\tiny{LCS}}$ \eqref{csaction} to ${\tilde S}_2$ \eqref{S2tilde} which 
brings the term $R$ in \eqref{f4}. Finally, one finds the metric field equation of EMG as \cite{Ozkan:2018cxj}:
\begin{equation} \label{emgeom}
\Lambda g_{\mu\nu} + G_{\mu\nu} -
\frac{1}{m^2} H_{\mu\nu} + \frac{1}{m^4} L_{\mu\nu} = 0\, , 
\end{equation}
where ($m$, $\Lambda$) are constants proportional to ($\mu$, $\Lambda_0$) and
\begin{eqnarray} \label{l}
    &&L_{\mu\nu} \equiv \frac{1}{2}e^{-1} \epsilon_{\mu}{}^{\rho\sigma}\epsilon_{\nu}{}^{\lambda
    \tau} C_{\rho\lambda}C_{\sigma\tau} \, ,\\ \label{h}
    && H_{\mu\nu} \equiv e^{-1} \epsilon_{(\mu|}{}^{\alpha\beta} \nabla_{\alpha} C_{\beta |\nu)}\,.
\end{eqnarray}

\section{Extended exotic massive gravity} \label{EMG2}
We can continue the game above and construct higher order exotic massive gravity theories. The rules of the game is to start from parity odd models, do the appropriate truncations and perhaps add some potential terms to make the final model third way consistent:
\begin{align}
    S_{2N+1}[f_{N+1}\to (f_N,\cdots,e)]\longrightarrow \tilde{S}_{2N}\, .
\end{align}
After showing how this construction works for the $N=0$ and $N=1$ cases, now we apply our method to the $N=2$ level. To do this, we start from $S_5$ constructed out of \eqref{parityodd} in
\cite{Afshar:2014ffa}
\begin{align} \label{l5}
    S_5=\frac{1}{\kappa^2 \mu^5}\int\left\langle e\cD f_3+f_1\cD f_2+\alpha_2h_2\left(R+ef_1\right)+h_1\left(\alpha_1 \cD h_1+f_1^2+ef_2\right)\right\rangle + S_3 \,,
\end{align}
where we used the rescaling freedoms to set some coefficients to unity. Starting from the $S_5$ action and making a general parity preserving truncation $f_3\to  (f_2,f_1,e)$, we may generate some terms which are already present in the $S_3$ \eqref{CSMG}
and $S_1$ \eqref{CSGaction} actions. 
Here, it turns out that only the $f_3\to e$ truncation gives rise to a new term.
We want to get a 
well-defined set of equations after shifting the connection with the remaining highest degree field, namely $h_2$. Therefore,
we also deform this action by adding appropriate potential terms and start with the following:
\begin{eqnarray} \label{emg2}
    \tilde{S}_{4} &= & S_3' + 
    \frac{1}{2\mu\kappa^2 }\int\left
    \langle\tfrac{2}{\mu^4}\left(\alpha_1 h_1\cD h_1 + \alpha_2h_2R \right)
    -\Lambda_0 e\cD e \right\rangle\\
    &+&  
    \frac{1}{\kappa^2 \mu^7} \int \left\langle  b_1ef_2h_2 + b_2 f_1^2h_2 + b_3h_1^3+ \tfrac{1}{\mu^2}b_4h_1^2h_2 + \tfrac{1}{\mu^{4}}b_5 h_2^2h_1 + \tfrac{1}{\mu^{6}}b_6 h_2^3   \right\rangle \nonumber
\end{eqnarray}
where we relaxed the coefficients in \eqref{CSMG} as 
\begin{equation}
    S_3' =\frac{1}{\kappa^2 \mu^3}\int \left( a_1e\cD f_2 + a_2 h_1R + a_3eh_1f_1 + \tfrac{\alpha_0}{2} f_1\cD f_1\right) \, .
\end{equation}
One can also add the $S_1$ action \eqref{CSGaction} 
to \eqref{emg2} but as we will see it is not needed to get a third-way consistent model. In \eqref{emg2} we have also ignored $f_2\cD f_1$
term coming from \eqref{l5} for the same reason.\footnote{\label{fdf}To be able to keep this term we need to add more potential terms to \eqref{emg2} and perform  a double shift $\omega \rightarrow {\omega} - c_1h_2 - c_2 h_1$.}

After we make the shift $\omega \rightarrow {\omega} - \frac{b_1}{a_1\mu^4}h_2$ and by
choosing
\begin{align}\label{3rdwayrelations}
\alpha_1    = -\frac{ b_1a_2^2}{4\alpha_2a_1} \, ,\quad
   b_2  = \frac{\alpha_0 b_1}{2a_1} \, ,\quad
   b_3    =  \frac{ b_1^2a_2^3}{12 \alpha_2^2a_1^2}  \, ,  \quad
   b_4 = \frac{ b_1^2a_2^2}{4 \alpha_2a_1^2}  \, , \quad
     b_5 = \frac{ b_1^2a_2^2}{2a_1^2} \, ,\quad
   b_6  =  \frac{\alpha_2 b_1^2}{6a_1^2} \, ,
\end{align}
we get the follwoing third-way consistent system: 
\begin{subequations}\label{emg2system}
\begin{align}
  \delta f_2\qquad  & {\cD} e = 0 \, , \label{first} \\
\delta h_2\qquad    &{R} +2 a_3\,[e,f_1] - \tfrac{\Lambda_0  }{2} [e,e] = 0 \, , \\
 \delta f_1\qquad   &  {\cD} f_1 +a_3\, [e,h_1] = 0 \, , \\
 \delta h_1\qquad   &{\cD} h_1 +  2  [e,f_2] +  [f_1,f_1] + \tfrac{a_3\mu^2}{2} [e,f_1] = 0 \, ,\\
 \delta e \qquad   & {\cD} f_2 +\tfrac{\Lambda_0}{\mu^2} [e,h_2] +  a_3\,[f_1,h_1] = 0 \, , \label{stop}\\
 \delta \omega \qquad   & \tfrac{1}{\mu^4}{\cD} h_2 -\tfrac{1}{\mu^8} [h_2,h_2] - \tfrac{1}{\mu^6} [h_2,h_1] -\tfrac{1}{4\mu^4} [h_1,h_1]  \nn\\ &- \tfrac{1}{2\mu^2} [f_1,f_1] - \tfrac{1}{\mu^2}[e,f_2]+2a_3[e,f_1] - \tfrac{\Lambda_0 }{2} [e,e] 
 = 0  \, \label{last} \, ,
\end{align}
\end{subequations}
where we used rescaling freedoms of the form fields to normalize non-zero free parameters $(a_1, a_2, \alpha_0, \alpha_2, b_1)$ to unity. Above  ${\cD}$ denotes the covariant derivative with respect to the shifted connection. Note that the cosmological constant $\Lambda_0$ and $a_3$ cannot  be set to zero as they are essential for solving the system consistently for $f_1$, $h_1$ and $h_2$. 
The full equation system \eqref{emg2system} resembles that of the extended NMG \cite{Afshar:2014ffa} deformed in the last equation with $h_2^2$ and $h_1h_2$ terms. But $f_1f_2$ term is missing since we did not include $f_2\cD f_1$ term in our action --- see the footnote \ref{fdf}. As noted above, we checked that adding 
$S_1$ action \eqref{CSGaction} to \eqref{emg2} is possible which does not change the general structure of the equations above but just modifies coefficients \eqref{3rdwayrelations}. 


To see the third-way consistency note that applying covariant derivative to equations \eqref{first}-\eqref{stop} and assuming  invertibility of the dreibein, we get constraints of the following sort
\begin{equation}
    e\cdot f_1= e\cdot h_1 = e\cdot f_2= e\cdot h_2  =0 \, .
\end{equation}
Now taking the covariant derivative of the 2-form on the left hand side of the equation \eqref{last}
and using these constraints we see that it vanishes only after we use the equation itself \eqref{last}:
\begin{equation}
    -\tfrac{2}{\mu^8} [\cD h_2,h_2] - \tfrac{1}{\mu^6} [\cD h_2,h_1] \approx 0 \, ,
\end{equation}
which shows that the model is third way consistent.


\subsection{Metric Field Equation}

To find the metric form of the equation system (\ref{first})-(\ref{last}) we first do the following shifts:

\begin{align}
    f_1  \rightarrow \hat{f}_1 + \frac{\Lambda_0}{4a_3} e \, , 
    \quad f_2  \rightarrow \hat{f}_2 - \left[\frac{\Lambda_0^2}{32a_3^2} + \frac{\mu^2 \Lambda_0}{16}\right]e \, . 
\end{align}
After these we get (we drop hats):
\begin{subequations}
\begin{align}
  \delta f_2\qquad  & {\cD} e = 0 \, ,  \\
\delta h_2\qquad    &{R} +2 a_3\,[e,f_1]  = 0 \, , \\
 \delta f_1\qquad   &  {\cD} f_1 +a_3\, [e,h_1] = 0 \, , \\
 \delta h_1\qquad   &{\cD} h_1 +  2  [e,f_2] +  [f_1,f_1] + \left[\tfrac{\Lambda_0}{2a_3} + \tfrac{a_3\mu^2}{2} \right] [e,f_1] = 0 \, ,\\
 \delta e \qquad   & {\cD} f_2 +\tfrac{\Lambda_0}{\mu^2} [e,h_2] +  a_3\,[f_1,h_1] + \tfrac{\Lambda_0}{4} [e,h_1] = 0 \, , \\
 \delta \omega \qquad   & \tfrac{1}{\mu^4}{\cD} h_2 -\tfrac{1}{\mu^8} [h_2,h_2] - \tfrac{1}{\mu^6} [h_2,h_1] -\tfrac{1}{4\mu^4} [h_1,h_1]  +  \tfrac{1}{2\mu^2} \cD h_1
 - \tfrac{9}{8} R + \tfrac{\Lambda_0 }{16} [e,e]
 = 0  \, . \label{fieldeqn}
\end{align}
\end{subequations}
The auxiliary fields can be solved as \cite{Afshar:2014ffa}:
\begin{align}
    (f_1)_{\mu\nu} & = - \frac{1}{2a_3}S_{\mu \nu} \, , \\
    (h_1)_{\mu\nu} & =  \frac{1}{2a_3^2} C_{\mu \nu}\, , \\
    (f_2)_{\mu\nu} & = - \frac{1}{4a_3^2} H_{\mu \nu} +  \frac{1}{4a_3^2}(P_{\mu \nu} - \frac{1}{4}P g_{\mu \nu}) 
    + \left[\frac{\Lambda_0}{8a_3^2} + \frac{\mu^2}{8} \right] S_{\mu \nu} \, , \\
     (h_2)_{\mu\nu} & = -\frac{\mu^2}{\Lambda_0} E_{\mu \nu} - \frac{\mu^2}{2\Lambda_0a_3^2}(Q_{\mu \nu} - \frac{1}{4}Q g_{\mu \nu}) + 
     \frac{\mu^2}{4\Lambda_0a_3^2}  SC_{\mu \nu} - \frac{\mu^2}{8a_3^2}C_{\mu \nu} \, ,
\end{align}
where the tensor $H$ is defined in \eqref{h} and $E,P$ and $Q$ tensors are given as
\begin{align}\label{DPEQten}
P_{\mu\nu}\equiv G_{\mu}{}^{\rho} S_{\nu \rho}\,, \qquad
E_{\mu\nu} \equiv e^{-1} \epsilon_{(\mu|}{}^{\alpha\beta} \nabla_{\alpha} f_{2\,\beta |\nu)}
\,,\qquad Q_{\mu\nu}\equiv C_{(\mu}{}^{\rho} S_{\nu) \rho}\, .
\end{align}
Here $G_{\mu \nu}$ is the Einstein, $S_{\mu \nu}$ is the Schouten \eqref{seqn}
and $C_{\mu \nu}$  is the Cotton tensor \eqref{ceqn}.  
We also define
\begin{eqnarray}
    &&X_{\mu\nu} \equiv e^{-1} \epsilon_{(\mu|}{}^{\alpha\beta} \nabla_{\alpha} h_{{2}_{\beta |\nu)}}\,, \\
    && Y_{\mu\nu} \equiv \frac{1}{2}e^{-1} \epsilon_{\mu}{}^{\rho\sigma}\epsilon_{\nu}{}^{\lambda
    \tau} h_{{2}_{\rho\lambda}}h_{{2}_{\sigma\tau}} \, , \\
    && Z_{\mu\nu} \equiv (h_{2})_{(\mu |\rho|} (h_{1})_{\nu)}{}^ {\rho} \, .
\end{eqnarray}
Now, from (\ref{fieldeqn}) one gets 
the following metric field equation of extended EMG as:
\begin{equation} \label{metriceom}
     \frac{1}{\mu^4} X_{\mu\nu} 
     -\frac{1}{\mu^8} Y_{\mu\nu} 
     - \frac{1}{\mu^6}  Z_{\mu\nu}
     -\frac{1}{16 a_3^4 \mu^4} L_{\mu\nu}  
     +\frac{1}{4a_3^2  \mu^2} H_{\mu\nu}
     - \frac{9}{8} G_{\mu\nu}
     + \frac{\Lambda_0 }{8}  g_{\mu\nu}
 = 0 \, ,
\end{equation}
where $L_{\mu\nu}$ is given in \eqref{l}.
Note that AdS is a solution if we identify 
$-\Lambda_0/9$ as the cosmological constant. Although, it seems that it would be possible to set $\Lambda_0=0$ at this level, recall that this is not allowed in the first order formulation.
Also notice that, the field equation (\ref{metriceom}) is 6th order in derivatives which can be viewed as a deformation  
of the EMG equation \eqref{emgeom} with 
$\{X, Y, Z\}$ tensors  and has one extra free parameter. Finally, one can also add the Cotton tensor $ßC_{\mu\nu}$ \eqref{ceqn} 
to the field equation \eqref{metriceom}, with a free coefficient by including the $S_1$ action \eqref{CSGaction} in \eqref{emg2}.

\section{Conclusion}\label{conc}
In this paper we showed how one may truncate a single  degree of freedom of the parity odd models $S_3$ and $S_5$ found in \cite{Afshar:2014ffa} to obtain exotic models $\tilde{S}_2$ and $\tilde{S}_4$ who have the same number of d.o.f. as their parity-even cousins $S_2$ and $S_4$. In these truncations { we first replaced the highest weight form fields in $S_3$ and $S_5$ with a linear combination of lower even-weight forms.} We then added sufficient extra potential terms, which violate the weight structure associated to the number of derivatives in the model, to make eventual models third way consistent. In the case of $\tilde{S}_2$, this extra term is given in \eqref{S2tilde} as $h_1^3$, and in the latter case ${\tilde S}_4$, these extra weight-violating interaction terms are given in \eqref{emg2} and can be generated from $(ef_2+f_1^2)h_2$ and $(h_1+\tfrac{1}{\mu^2}h_2)^3$ generating functions. Models $\tilde{S}_2$ and $\tilde{S}_4$ are third way consistent after shifting the spin-connection $\omega$ with $h_1$ and $h_2$ respectively.\footnote{The case of $S_1\to{\tilde S}_0$ is special as the number of degrees of freedom do not change and we do not add any new term to the Lagrangian. This is because $L_1$ has an extra conformal gauge symmetry \cite{Afshar:2011qw}.} 
{In general, for constructing exotic models ${\tilde S}_{2N}$ our method  has three main steps;
\begin{enumerate}
    \item The field $f_{N+1}$ in $ S_{2N+1}$ which has the highest even-weight $2N+2$ is identified with a linear combination of  $f_N,\, f_{N-1},\cdots$ which fixes dynamical terms in ${\tilde S}_{2N}$.
    \item Having fixed the dynamical terms, necessary potential terms are easy to figure out by requiring the equation system to be third-way solvable. In particular, $h_N^3$ should be present.  
    \item The spin connection is shifted $\omega\to\omega +h_{N}+(h_{N-1}+\cdots)$ so that there is no torsion w.r.t. the new connection.
\end{enumerate}}
 By restricting both dynamical and potential terms, our method provides a systematic way of constructing infinitely many exotic third way consistent models. This approach may also be helpful in obtaining supersymmetric versions of these models which has not been achieved until now.

In \cite{Afshar:2014dta} it is shown that there exists a scaling limit, or a flow, from certain interacting multi-gravity theories; $S_0[e_0,\omega_0]+\cdots+S_0[e_N,\omega_N] + (\text{appropriate interaction terms})$ to $S_{2N}$ models. Here we expect a same pattern with ${\tilde S}_0[e_0,\omega_0]+\cdots+{\tilde S}_0[e_N,\omega_N] + (\text{appropriate interaction terms})$ to hold for ${\tilde S}_{2N}$ as well. This has been shown to work for the $N=1$ case, i.e. EMG  \cite{Ozkan:2018cxj}, already in \cite{Ozkan:2019iga}.

It is desirable to study the physical properties of the extended EMG model \eqref{metriceom}
that we constructed in this paper further. For instance, it would be interesting to study its matter couplings \cite{Arvanitakis:2014yja, Ozkan:2018cxj} as well as its
unitary extensions \cite{Ozkan:2019iga}, which is hard to achieve 
in such models \cite{Alkac:2017vgg}.
For the EMG \cite{Ozkan:2018cxj}, the
existence of asymptotically AdS solutions obeying different boundary conditions was found in \cite{Chernicoff:2018hpb, Giribet:2019vbj}. 
Investigating this problem for our model \eqref{metriceom} would be important for understanding effects of higher derivative terms on energy \cite{Mann:2018vum} and causality \cite{Kilicarslan:2019ply}. 

A generic feature in these models is that field equations in the metric form have opposite parity compared to the (first order) action. This is a crucial point that should be taken into account for computing charges. Obviously this leads to different conclusions depending on whether one uses on-shell approaches like ADT \cite{Mann:2018vum}   
or off-shell ones for this computation,
see \cite{Adami:2017phg}) for a review and references. Such a discrepancy is generic for third way consistent models \cite{Tekin:2014jna,Bergshoeff:2018luo}. In exotic models, for the BTZ black hole solution, one expects the role of mass and angular momentum to be exchanged in these two approaches:\footnote{{ This claim was later proved in \cite{Bergshoeff:2019rdb}.}}
\begin{align}
M_{\text{\tiny on-shell}}=J_{\text{\tiny off-shell}}\,,\qquad\qquad J_{\text{\tiny on-shell}}=M_{\text{\tiny off-shell}}\,.
\end{align}
Moreover, in the off-shell approach the left and right central charges have opposite signs while in the on-shell calculation they have the same sign. Their absolute values obtained in both approaches are of course the same. This exchange between charges and their signs has been studied in detail for the 3D Einstein gravity $S_0$ \eqref{einstein}
and 3D exotic gravity ${\tilde S}_0$ \eqref{exoticEH}
in \cite{Afshar:2011qw,Townsend:2013ela}, see also \cite{Park:2006hu}. We expect this to be a general feature of charges of higher exotic models ${\tilde S}_{2N}$ compared to $S_{2N}$.

A natural question to ask is what happens if we continue truncating further after getting an exotic model. In appendix \ref{mmgemg} we show that truncating the highest degree field in EMG \cite{Ozkan:2018cxj} in a parity violating way
one obtains MMG \cite{Bergshoeff:2014pca} . We expect this to hold for higher order exotic models as well: 
\begin{align}
    &{S}_{\text{\tiny EMG}}^{\text{\tiny (N)}} \longrightarrow S_{\text{\tiny MMG}}^{\text{\tiny (N)}} \, .
\end{align}
\noindent For example truncating the $h_2$ field in our extended EMG model \eqref{metriceom} as 
$h_2 \rightarrow (c_1f_2 + c_2f_1 + c_3 e)$ we anticipate 
to obtain extended MMG with 5th order derivatives. Another interesting open problem is to see whether the third way consistent models obtained in \cite{Geiller:2018ain, Geiller:2019dpc} can fit into this scheme.
Finally, finding applications of such models in condensed matter physics \cite{Grumiller:2013at}
would be nice. We leave investigation of these connections to a future work.






\section*{Acknowledgements}
We would like to thank Daniel Grumiller, Jan Rosseel and Shahin Sheikh-Jabbari for useful discussions. NSD is partially supported by the Scientific and Technological
Research Council of Turkey (T\"ubitak) Grant No.116F137. Authors wish to thank each others institutions namely IPM, Tehran and Bo\u gazi\c ci University for hospitality where this work was initiated and developed during their mutual visits and gratefully acknowledge financial support of ICTP network scheme NT-04 for these. NSD is grateful to ESI, Vienna for hospitality and financial support where this work was completed while he was visiting for the Research in Teams Programme.

\appendix

\section{MMG from EMG by Truncation} \label{mmgemg}
Minimal massive gravity \eqref{mmgaction}
has 3 dynamical fields ($e,\omega,f_1$). Here we ask if we can obtain it as a truncation from a four field model. At level 4 there are two available models namely
NMG \cite{Bergshoeff:2009hq, Bergshoeff:2009aq} and EMG \cite{Ozkan:2018cxj}. It is easy to see that it is not possible to get MMG from NMG via truncation. If instead we start with EMG \eqref{CSMG} and make the parity violating truncation $h_1 \rightarrow (c_4f_1 + c_5 e)$ we get the following generic action; 
\begin{equation}
    S = S_{LCS} + \int \left\langle c_0f_1 \cD e + c_1 e\cD e +  c_2 ef_1^2 + c_3f_1^3 + c_4 Rf_1  + c_5 eR + c_6 e^2 f_1 + c_7 e^3 \right\rangle \, .
\end{equation}
Here we should treat all constants independent since some coefficients in the initial action \eqref{CSMG} are set to one by re-scaling fields 
which may be spoiled now. Field equations are
\begin{align}
    \delta f_1 \qquad  & c_0 \cD e + 2c_2 [e,f_1] + 3c_3[f_1,f_1] + c_4R + c_6[e,e] = 0  \, , \\
    \delta \omega \qquad  & R + c_0[e,f_1] + c_1 [e,e] + c_4\cD f_1 + c_5 \cD e = 0 \, , \\
    \delta e \qquad  & c_0\cD f_1 + 2c_1\cD e + c_2[f_1,f_1] + c_5 R + 2c_6 [e,f_1] + 3c_7 [e,e] =0 \, .
\end{align}
If we now demand that the field equations after the shift $\omega \rightarrow \omega - \alpha f_1$
take the form of MMG$_1$
\begin{align}
   & {\cD} e = 0 \, , \\
    & {R} + m[e,f_1] + k [e,e] = 0 \, , \\
    & {\cD}f_1 + n [f_1,f_1] + p[e,f_1] + q[e,e] = 0 \, ,
\end{align}
one gets an under-determined system. If $c_4 = 0$, this implies 
$c_3=c_6=0$ and the non-zero coefficients are found to be:
\begin{equation}
    c_2/c_0^2=-n\equiv \frac{\alpha}{2} \, ,  \,   \, c_7=\frac{q + c_5 c_1}{3}  \, , k=2(c_1+\alpha q)  \, ,
    \, m=2(1-c_5\alpha)^2  \, ,  \, p= c_5^2\alpha -c_5 \, .   
\end{equation}
When $c_1=0$, these are exactly the coefficients of MMG \cite{Bergshoeff:2014pca} with $c_5=-\sigma$
and $c_7=\Lambda_0/6$. If $c_4 \neq 0$, then one can shift fields to obtain the same model.

\bibliographystyle{fullsort.bst}
 
\bibliography{references} 

\end{document}